\documentclass[10pt]{article}
\usepackage[dvips]{color}
\usepackage{epsfig}
\usepackage{amsmath}
\usepackage{graphicx}
\usepackage{amssymb}
\usepackage{float}

\begin{document}
\title{\bf Exact Calculation of the Capacitance and the Electrostatic
Potential Energy for a Nonlinear Parallel-Plate Capacitor in a
Two-Parameter Modification of Born-Infeld Electrodynamics}

\author{S. K. Moayedi \thanks{Corresponding author, E-mail:
s-moayedi@araku.ac.ir}\hspace{1mm} , F. Fathi
\thanks{E-mail: f-fathi@phd.araku.ac.ir}\hspace{1mm}\\
{\small {\em  Department of Physics, Faculty of Sciences,
Arak University, Arak 38156-8-8349, Iran}}\\
}
\date{\small{}}
\maketitle

\begin{abstract}
\noindent The nonlinear capacitors are important devices in modern
technologies and applied physics. The aim of this paper is to
calculate exactly the capacitance and the electrostatic potential
energy of a nonlinear parallel-plate capacitor by using a
two-parameter modification of Born-Infeld
electrodynamics. Our
calculations show that the capacitance and the electrostatic
potential energy of a nonlinear parallel-plate capacitor in
modified Born-Infeld theory have the weak field expansions
$C=\frac{\epsilon_{0} A}{d}+{\cal O}(q^{2})$ and
$U=\frac{q^{2}}{2(\frac{\epsilon_{0} A}{d})}+{\cal O}(q^{4})$,
where $q$ is the amount of electric charge on each plate of the
capacitor. It is demonstrated that the results of this paper are
in agreement with the results of Maxwell electrodynamics for weak
electric fields. Numerical evaluations show that the nonlinear
electrodynamical effects in modified Born-Infeld theory are
negligible in the weak field regime.

\noindent
\hspace{0.35cm}



\end{abstract}

\section{Introduction}
The nonlinear capacitors have found a wide range of applications
in circuit theory in electrical engineering and different branches
of applied physics [1-7]. In a nonlinear capacitor, in contrast
with the ordinary capacitors, the capacitance is a function of
voltage, i.e., $C=f(\triangle \phi)$ where $\triangle \phi$ is the
potential difference between the plates of the capacitor and the
function $f(\triangle \phi)$ can be determined empirically or from
a fundamental electromagnetic theory like Maxwell electrodynamics
or Born-Infeld theory [1,2,8-10]. In Ref. [2], it has been shown
that for a nonlinear capacitor which its capacitance depends
linearly on the voltage, the usual relation
$U=\frac{1}{2}C(\triangle \phi)^{2}$ is not satisfied. The authors
of Ref. [6] have shown that the electrostatic potential energy of
a nonlinear capacitor can be expanded as follows:
\begin{equation}
U=\alpha q^{2}+\beta q^{4}+\gamma q^{6}+...,
\end{equation}
where $\alpha$, $\beta$, $\gamma$, and $...$ are material
dependent constants. Today we know that the interaction between
the charged bodies can be described by Maxwell equations
classically [8]. On the other hand, Maxwell electrodynamics
suffers from serious difficulties such as infinite self-energy of
the point charges [8]. In 1934 Born and Infeld introduced the
following Lagrangian density\footnote{We use SI units in this
paper. The space-time metric has the signature $(+,-,-,-)$.} [9]:
\begin{equation}
{\cal L}_{_{BI}}=\epsilon_{0}\beta^{2}\bigg\lbrace
1-\sqrt{1+\frac{c^{2}}{2\beta^{2}}F_{\mu\nu}F^{\mu\nu}-\frac{c^{4}}{16\beta^{4}}(F_{\mu\nu}\,\star
F^{\mu\nu})^{2}}\bigg\rbrace,
\end{equation}
where $F_{\mu\nu}=\partial_{\mu}A_{\nu}-\partial_{\nu}A_{\mu}$ is
the Faraday tensor, $A^{\mu}=(\frac{1}{c} \phi , \textbf A)$ is
the gauge potential, $\star
F^{\mu\nu}=\frac{1}{2}\epsilon^{\mu\nu\alpha\beta}F_{\alpha\beta}$
is the dual field tensor, and $\beta$ is the maximum value of the
electric field in Born-Infeld theory. In string theory the
dynamics of electromagnetic fields on $D$-branes can be
represented by a Born-Infeld type theory [10]. The solutions of
Born-Infeld equations for an infinite charged line and an
infinitely long cylinder have been obtained in Ref. [11]. In
1930s, Heisenberg, Euler, and Kockel studied the scattering of
light by light according to Dirac's hole theory [12-14]. They
showed that Maxwell electrodynamics should be corrected by adding
nonlinear terms due to the quantum electrodynamical effects
[12-14]. It must be emphasized that, for weak electromagnetic
fields, the Lagrangian density and the energy density of nonlinear
electrodynamics have the following explicit expressions:
\begin{equation}
{\cal
L}=\sum_{i=0}^{\infty}\sum_{j=0}^{\infty}c_{_{i,j}}F^{i}G^{j},
\end{equation}
\begin{equation}
u=\sum_{i=0}^{\infty}\sum_{j=0}^{\infty}c_{_{i,j}}\big(2\epsilon_{0}
i F^{(i-1)}G^{j}\textbf E^{2}+(j-1)F^{i}G^{j}\big),
\end{equation}
where $c_{_{i,j}}$ are field-independent parameters,
$F:=\big(\epsilon_{0}\textbf
{E}^{2}-\frac{\textbf{B}^{2}}{\mu_{0}}\big)$, and
$G:=\sqrt{\frac{\epsilon_{0}}{\mu_{0}}}(\textbf{E}.\textbf{B})$
[15,16]. Note that in the weak field regime (2) is a particular
example of (3). The effect of nonlinear corrections on the electric
field between the plates of a parallel-plate capacitor has been
studied in the framework of Heisenberg-Euler-Kockel electrostatics
[17]. In a recent paper, the capacitance and the electrostatic
potential energy for a parallel-plate and spherical capacitors have
been computed in ordinary Born-Infeld theory [18]. Iacopini and
Zavattini suggested and developed a $(p,\tau)$-two-parameter
modification of Born-Ifeld electrodynamics, in which the
electrostatic self-energy of a point charge becomes a finite value
for $p<1$ [19].  The most important aim of this paper is to
calculate the capacitance of a nonlinear parallel-plate capacitor
from the viewpoint of Iacopini-Zavattini modification of Born-Infeld
electrodynamics [19]. Another aim is to show that the nonlinear
phenomena in electrodynamics are negligible for weak electric
fields. This paper is organized as follows. In Section 2, the
formulation of modified Born-Infeld electrodynamics coupled to an
external current is presented. In Section 3, the symmetric
energy-momentum tensor for modified Born-Infeld electrodynamics is
constructed from the canonical energy-momentum tensor by using
Belinfante's procedure. In Section 4, we show that the capacitance
and the electrostatic potential energy for a nonlinear
parallel-plate capacitor in modified Iacopini-Zavattini
electrodynamics can be calculated exactly when the parameter $p$
takes the values
$\lbrace\frac{1}{2},\frac{2}{3},\frac{3}{4},\frac{5}{6}\rbrace$. It
is verified that the results of Section 4 for $p=\frac{1}{2}$ are
compatible with those obtained previously in [18]. Numerical
evaluations in summary and conclusions indicate that the nonlinear
corrections to the electrostatic potential energy of a
parallel-plate capacitor in modified Born-Infeld electrostatics are
not important in the weak field regime.

\section{A Brief Review of Modified Born-Infeld Electrodynamics}
The modified Born-Infeld electrodynamics in a (3+1)-dimensional
Minkowski space-time is described by the following Lagrangian
density (see Eq. (B.10) in Ref. [19]):
\begin{equation}
{\cal L}_{_{p,\tau}}=\frac{1}{2p}\epsilon_{0}\beta^{2}\bigg\lbrace
1-\bigg[1+\frac{c^{2}}{2\beta^{2}}F_{\mu\nu}F^{\mu\nu}-\tau\frac{c^4}{16\beta^{4}}\big(F_{\mu\nu}\,\star
F^{\mu\nu}\big)^{2}\bigg]^{p}\bigg\rbrace -J^{\mu}A_{\mu},
\end{equation}
where $p<1$ is a real dimensionless constant, $\tau$ is another
dimensionless constant, and $J^{\mu}=(c \rho,\textbf J)$ is an
external current for the $U(1)$ gauge field $A^{\mu}$. The
parameter $\beta$ in Eq. (5) is called the Born-Infeld parameter
and shows the upper limit of the electric field in modified
Born-Infeld electrodynamics. It is necessary to note that for
$p=\frac{1}{2}$, $\tau=1$ the Lagrangian density in Eq. (5)
becomes the standard Born-Infeld Lagrangian density [9], while for
$p=1$, $\tau=0$ we obtain the Maxwell Lagrangian density. The
equation of motion for the vector field $A_{\lambda}$ is
\begin{equation}
\frac{\partial{\cal L}_{_{p,\tau}}}{\partial
A_{\lambda}}-\partial_{\sigma}\bigg(\frac{\partial{\cal
L}_{_{p,\tau}}}{\partial(\partial_{\sigma}A_{\lambda})}\bigg)=0.
\end{equation}
If we put Eq. (5) into Eq. (6), we will get the inhomogeneous
modified Born-Infeld equations as follows:
\begin{equation}
\partial_{\sigma}\bigg(\frac{F^{\sigma\lambda}-\tau\frac{c^{2}}{4\beta^{2}}\big(F_{\mu\nu} \star F^{\mu\nu}\big)\star F^{\sigma\lambda}}{\big[1+\frac{c^{2}}{2\beta^{2}}F_{\mu\nu}F^{\mu\nu}-\tau\frac{c^4}{16\beta^4}\big(F_{\mu\nu} \star F^{\mu\nu}\big)^{2}\big]^{1-p}}\bigg)=\mu_{0}J^{\lambda}.
\end{equation}
The dual field tensor $\star F^{\mu\nu}$ satisfies the following
Bianchi identity:
\begin{equation}
\partial_{\mu}\star F^{\mu\nu}=0.
\end{equation}
In (3+1)-dimensional space-time, the Faraday 2-form $F$ and its
dual $\star F$ have the following expressions [20]:
\begin{eqnarray}
F&=&F_{0i}\;dx^{0}\wedge dx^{i}+\frac{1}{2} F_{ij}\;dx^{i} \wedge
dx^{j}\nonumber\\
&=& E^{i}\;dt\wedge dx^{i}-\frac{1}{2}\epsilon^{ijk}
B^{i}\;dx^{j}\wedge dx^{k},\\
\star F&=&-B^{i}\;dt\wedge
dx^{i}-\frac{1}{2}\epsilon^{ijk}\;E^{i}\;dx^{j}\wedge dx^{k},
\end{eqnarray}
where $i,j,k=1,2,3$, and
$$ \lbrace E^{i} \rbrace=\lbrace E_{x},E_{y},E_{z} \rbrace, \;\; \lbrace B^{i}\rbrace=\lbrace B_{x},B_{y},B_{z} \rbrace. $$
Using Eqs. (9) and (10), Eqs. (7) and (8) take the following
vector forms:
\begin{eqnarray}
\boldsymbol{\nabla} \cdot \textbf D(\textbf x ,t) &=& \rho(\textbf x,t), \\
 \boldsymbol{\nabla} \times \textbf H(\textbf x ,t) &=& \textbf J(\textbf x ,t)+\frac{\partial
\textbf D(\textbf x ,t)}{\partial t}, \\
\boldsymbol{\nabla} \cdot \textbf B (\textbf x ,t) &=& 0, \\
 \boldsymbol{\nabla} \times \textbf E(\textbf x ,t) &=& -\frac{\partial
\textbf B(\textbf x ,t)}{\partial t},
\end{eqnarray}
where $\textbf D(\textbf x ,t)$ and $\textbf H(\textbf x ,t)$ are
given by
\begin{eqnarray}
\textbf D(\textbf x ,t) &=& \epsilon_{0}\frac {\textbf E(\textbf x ,t)+\tau\frac{c^{2}}{\beta^{2}}\big(\textbf E(\textbf x ,t).\textbf B(\textbf x ,t)\big)\textbf B(\textbf x ,t)}{\Omega_{_{p,\tau}}\;^{^{\frac{1}{p}-1}}}, \\
 \textbf H(\textbf x ,t) &=& \frac{1}{\mu_{0}}\frac {\textbf B(\textbf x ,t)-\tau\frac{1}{\beta^{2}}\big(\textbf E(\textbf x ,t).\textbf B(\textbf x ,t)\big)\textbf E(\textbf x ,t)}{\Omega_{_{p,\tau}}\;^{^{\frac{1}{p}-1}}},
\end{eqnarray}
and $\Omega_{_{p,\tau}}$ is defined as follows:
\begin{equation}
\Omega_{_{p,\tau}}:=\bigg[1+\frac{c^{2}}{2\beta^{2}}F_{\mu\nu}F^{\mu\nu}-\tau\frac{c^{4}}{16\beta^{4}}\big(F_{\mu\nu}
\star F^{\mu\nu}\big)^{2}\bigg]^{p}.
\end{equation}
Now, let us study the electrostatic case where $\textbf B =\textbf
J =0$ and all other physical quantities are time independent. In
this case the modified Born-Infeld equations (11)-(14) are
\begin{eqnarray}
\boldsymbol{\nabla} \cdot\bigg(\frac{\textbf E(\textbf
x)}{[1-\frac{\textbf E^{2}(\textbf x)}{\beta^{2}}]^{^{1-p}}}\bigg)
&=&
\frac{\rho(\textbf x )}{\epsilon_{0}},  \\
\boldsymbol{\nabla} \times \textbf E (\textbf x ) &=& 0 .
\end{eqnarray}
The above equations are basic equations of modified Born-Infeld
electrostatics [19]. By using the divergence theorem in vector
calculus, we get the integral form of Eq. (18) as follows:
\begin{equation}
\oint_{C_{_{2}}}\frac{1}{[1-\frac{\textbf E^{2}(\textbf
x)}{\beta^{2}}]^{^{1-p}}}\;\textbf E(\textbf
x).{\hat{\textbf{n}}}\; da =\frac{1}{\epsilon_{0}} \int_{C_{_{3}}}
\rho (\textbf x) d^{3}x ,
\end{equation}
where  $C_{_{2}}$ is a $2$-chain which is the boundary of a
$3$-chain $C_{_{3}}$, i.e., $C_{_{2}}=\partial C_{_{3}}$ [20].
Equation (20) is Gauss's law in modified Born-Infeld
electrostatics.

\section{The Symmetric Energy-Momentum Tensor for Modified \\
Born-Infeld Electrodynamics} In this section, we obtain the
symmetric energy-momentum tensor for modified Born-Infeld
electrodynamics. According to Eq. (5), the Lagrangian density for
modified Born-Infeld electrodynamics in the absence of external
current $J^\mu$ is
\begin{equation}
{\cal L}_{_{p,\tau}}=\frac{1}{2p}\epsilon_{0}\beta^{2}\bigg\lbrace
1-\bigg[1+\frac{c^{2}}{2\beta^{2}}F_{\mu\nu}F^{\mu\nu}-\tau\frac{c^4}{16\beta^{4}}\big(F_{\mu\nu}\,\star
F^{\mu\nu}\big)^{2}\bigg]^{p}\bigg\rbrace.
\end{equation}
From (21), we derive the following classical field equation:
\begin{equation}
\partial_{\sigma}\bigg(\frac{F^{\sigma\lambda}-\tau\frac{c^{2}}{4\beta^{2}}\big(F_{\mu\nu} \star F^{\mu\nu}\big)\star F^{\sigma\lambda}}{\Omega_{_{p,\tau}}\;^{^{\frac{1}{p}-1}}}\bigg)=0.
\end{equation}
The canonical energy-momentum tensor for Eq. (21) is [21-23]
\begin{equation}
\Theta^{^{\sigma}}\;_{_{\eta}}=\frac{\partial{\cal
L}_{_{p,\tau}}}{\partial(\partial_{\sigma}
A_{\lambda})}(\partial_{\eta}
A_{\lambda})-\delta^{^{\sigma}}\;_{_{\eta}}\;{\cal L}_{_{p,\tau}}.
\end{equation}
If we substitute (21) into (23) and use (22), we will obtain the
following expression for the canonical energy-momentum tensor
$\Theta^{^{\sigma}}\;_{_{\eta}}$:
\begin{equation}
\Theta^{^{\sigma}}\;_{_{\eta}}=\frac{1}{\mu_{0}}\frac{F^{\sigma\lambda}-\tau\frac{c^{2}}{4\beta^{2}}\big(F_{\mu\nu}
\star F^{\mu\nu}\big)\star
F^{\sigma\lambda}}{\Omega_{_{p,\tau}}\;^{^{\frac{1}{p}-1}}}F_{\lambda\eta}+\frac{1}{2p}\epsilon_{0}\beta^{2}(\Omega_{_{p,\tau}}\;-1)\;\delta^{^{\sigma}}\;_{_{\eta}}+\partial_{\lambda}\;{\cal
R}^{^{\lambda\sigma}}\;_{_{\eta}},
\end{equation}
where
\begin{eqnarray}
{\cal
R}^{^{\lambda\sigma}}\;_{_{\eta}}&:=&\frac{1}{\mu_{0}}\frac{F^{\lambda\sigma}-\tau\frac{c^{2}}{4\beta^{2}}\big(F_{\mu\nu}
\star F^{\mu\nu}\big)\star F^{\lambda\sigma}}{\Omega_{_{p,\tau}}\;^{^{\frac{1}{p}-1}}}A\;_{\eta}, \\
{\cal R}^{^{\sigma\lambda}}\;_{_{\eta}}&=&-{\cal
R}^{^{\lambda\sigma}}\;_{_{\eta}}.
\end{eqnarray}
It is well known that the canonical energy-momentum tensor
$\Theta^{^{\sigma}}\;_{_{\eta}}$ in (23) is generally not
symmetric [21-24]. Belinfante showed that the canonical
energy-momentum tensor $\Theta^{^{\sigma}}\;_{_{\eta}}$ in Eq.
(23) can be written as follows [21]:
\begin{equation}
\Theta^{^{\sigma}}\;_{_{\eta}}=T^{^{\sigma}}\;_{_{\eta}}+\partial_{\lambda}\;{\cal
R}^{^{\lambda\sigma}}\;_{_{\eta}},
\end{equation}
where the second- and third-order tensors
$T^{^{\sigma}}\;_{_{\eta}}$ and ${\cal
R}^{^{\lambda\sigma}}\;_{_{\eta}}$ must satisfy the following
conditions:
\begin{subequations}
\begin{eqnarray}
T^{^{\sigma\eta}}&=&T^{^{\eta\sigma}}, \\
{\cal R}^{^{\lambda\sigma}}\;_{_{\eta}}&=&-{\cal
R}^{^{\sigma\lambda}}\;_{_{\eta}}.
\end{eqnarray}
\end{subequations}
The second-order tensor $T^{^{\sigma\eta}}$ in the above equations
is called the symmetric energy-momentum tensor [22]. A comparison
between Eqs. (24) and (27) clearly shows that the symmetric
energy-momentum tensor for modified Born-Infeld electrodynamics is
\begin{equation}
T^{^{\sigma}}\;_{_{\eta}}=\frac{1}{\mu_{0}}\frac{F^{\sigma\lambda}-\tau\frac{c^{2}}{4\beta^{2}}\big(F_{\mu\nu}
\star F^{\mu\nu}\big)\star
F^{\sigma\lambda}}{\Omega_{_{p,\tau}}\;^{^{\frac{1}{p}-1}}}F_{\lambda\eta}+\frac{1}{2p}\epsilon_{0}\beta^{2}(\Omega_{_{p,\tau}}\;-1)\;\delta^{^{\sigma}}\;_{_{\eta}}.
\end{equation}
After straightforward but tedious calculations, one finds that in
the presence of an external current the symmetric energy-momentum
tensor $T^{^{\sigma}}\;_{_{\eta}}$ in (29) satisfies the following
equation:\footnote{It is obvious that for source-free modified
Born-Infeld theory the right-hand side of (30) vanishes, i.e.,
$\partial_{\sigma}\;T^{^{\sigma}}\;_{_{\eta}}=0$.}
\begin{equation}
\partial_{\sigma}\;T^{^{\sigma}}\;_{_{\eta}}=J^{^{\sigma}}F_{\sigma\eta}.
\end{equation}
Using Eqs. (9) and (10) together with Eq. (29), the energy density
of modified Born-Infeld electrodynamics is given by
\begin{eqnarray}
u(\textbf x,t)&=&T^{0}\;_{0}(\textbf x,t)\nonumber\\
 &=&\frac{1}{2p}\epsilon_{0}\beta^{2}
\bigg\lbrace\frac{(2p-1)\bigg(\frac{\textbf E^{2}(\textbf
x,t)}{\beta^{2}}+\tau\frac{\big(\textbf E(\textbf x,t).c\;\textbf
B(\textbf x,t)\big)^{2}}{\beta^{4}}\bigg)+\frac{c^{2}\textbf
B^{2}(\textbf x,t)}{\beta^{2}}+1}{\bigg[1-\frac{\big(\textbf
E^{2}(\textbf x,t)-c^{2}\textbf B^{2}(\textbf
x,t)\big)}{\beta^{2}}-\tau\frac{\big(\textbf E(\textbf
x,t).c\;\textbf B(\textbf
x,t)\big)^{2}}{\beta^{4}}\bigg]^{^{1-p}}}-1\bigg\rbrace.
\end{eqnarray}
According to Eq. (31), the energy density of an electrostatic
field in modified Born-Infeld electrodynamics becomes
\begin{equation}
u(\textbf x )=T^{0}\;_{0}(\textbf x
)=\frac{1}{2p}\epsilon_{0}\beta^{2}
\bigg[\frac{(2p-1)\frac{\textbf E^{2}(\textbf
x)}{\beta^{2}}+1}{\big(1-\frac{\textbf E^{2}(\textbf
x)}{\beta^{2}}\big)^{^{1-p}}}-1\bigg].
\end{equation}
For $p=\frac{1}{2}$, the modified electrostatic energy density in
Eq. (32) becomes the energy density of an electrostatic field in
Born-Infeld electrodynamics, i.e.,
\begin{equation}
u(\textbf x )=\epsilon_{0}\beta^{2}
\bigg(\frac{1}{\sqrt{1-\frac{\textbf E^{2}(\textbf
x)}{\beta^{2}}}}-1\bigg).
\end{equation}

\section{Calculation of the Capacitance and the Electrostatic Potential Energy of a Nonlinear Parallel-Plate Capacitor in Modified Born-Infeld Theory}
In order to calculate the capacitance of a nonlinear
parallel-plate capacitor in modified Born-Infeld electrostatics,
we assume a capacitor composed of two large parallel conducting
plates with area $A$ and separation $d$ (see Figure 1).
\begin{figure}[ht]
\centerline{\includegraphics[width=6.4 cm]{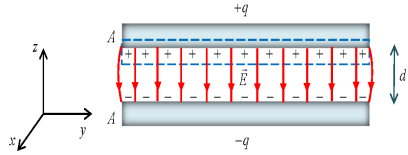}} \caption{\small
A parallel-plate capacitor. The Gaussian surface is represented by
dashed lines. The symmetry of the problem implies that
$\textbf{E}(\textbf{x})=E_{z}(-\hat{\textbf{e}}_{z})$, where
$\,\hat{\textbf{e}}_{z}$ is the unit vector in the $z$-direction.}
\end{figure}
\\
By applying modified Gauss's law  in (20) to the Gaussian surface
in Figure 1, we obtain the following equation for $E_{z}$
\begin{equation}
E_{z}\;^{^{\Gamma}}+\bigg(\frac{q}{\beta^{\frac{2}{\Gamma}}\epsilon_{0}A}
\bigg)^{^{\Gamma}}\; E_{z}\;^{^{2}}-\bigg(\frac{q}{\epsilon_{0}A}
\bigg)^{^{\Gamma}}=0,
\end{equation}
where $\Gamma:=\frac{1}{1-p}$. Now, let us obtain the exact
solutions of Eq. (34) for
$p\in\lbrace\frac{1}{2},\frac{2}{3},\frac{3}{4},\frac{5}{6}\rbrace$.
\\
For $p=\frac{1}{2}(\Gamma=2)$, Eq. (34) becomes the quadratic
equation
\begin{equation}
\bigg[1+\bigg(\frac{q}{\beta\epsilon_{0}A}
\bigg)^{^{2}}\bigg]E_{z}\;^{^{2}}-\;\bigg(\frac{q}{\epsilon_{0}A}
\bigg)^{^{2}}=0. \nonumber
\end{equation}
For $p=\frac{2}{3}(\Gamma=3)$, Eq. (34) becomes the cubic equation
\begin{equation}
E_{z}\;^{^{3}}+\bigg(\frac{q}{\beta^{\frac{2}{3}}\epsilon_{0}A}
\bigg)^{^{3}}\; E_{z}\;^{^{2}}-\bigg(\frac{q}{\epsilon_{0}A}
\bigg)^{^{3}}=0. \nonumber
\end{equation}
For $p=\frac{3}{4}(\Gamma=4)$, Eq. (34) becomes the following
quartic equation
\begin{equation}
E_{z}\;^{^{4}}+\bigg(\frac{q}{\beta^{\frac{1}{2}}\epsilon_{0}A}
\bigg)^{^{4}}\; E_{z}\;^{^{2}}-\bigg(\frac{q}{\epsilon_{0}A}
\bigg)^{^{4}}=0. \nonumber
\end{equation}
Finally, for $p=\frac{5}{6}(\Gamma=6)$, (34) becomes the following
sextic equation
\begin{equation}
E_{z}\;^{^{6}}+\bigg(\frac{q}{\beta^{\frac{1}{3}}\epsilon_{0}A}
\bigg)^{^{6}}\; E_{z}\;^{^{2}}-\bigg(\frac{q}{\epsilon_{0}A}
\bigg)^{^{6}}=0. \nonumber
\end{equation}
Note that the above sextic equation by a suitable change of variable
reduces to a cubic equation. In Galois theory, the Abel-Ruffini
theorem or Abel's impossibility theorem says that: \textit{for all
$n\geq5$, there is a polynomial in $\mathbb{Q}[x]$ of degree $n$
that is not solvable by irreducible radicals over $\mathbb{Q}$}
[25,26].\footnote{$\mathbb{Q}$ is the field of rational numbers (see
page 116 in Ref. [26]).} According to Abel's impossibility theorem,
equation (34) has the following exact solutions for
$p\in\lbrace\frac{1}{2},\frac{2}{3},\frac{3}{4},\frac{5}{6}\rbrace$
\begin{subequations}
\begin{eqnarray}
E_{z}\;^{^{p=\frac{1}{2}}}&=&\frac{q}{\epsilon_{0}A}\Pi\;^{^{p=\frac{1}{2}}}(q),\\
E_{z}\;^{^{p=\frac{2}{3}}}&=&\frac{q}{\epsilon_{0}A}\Pi\;^{^{p=\frac{2}{3}}}(q),\\
E_{z}\;^{^{p=\frac{3}{4}}}&=&\frac{q}{\epsilon_{0}A}\Pi\;^{^{p=\frac{3}{4}}}(q),\\
E_{z}\;^{^{p=\frac{5}{6}}}&=&\frac{q}{\epsilon_{0}A}\Pi\;^{^{p=\frac{5}{6}}}(q),
\end{eqnarray}
\end{subequations}
where
\begin{subequations}
\begin{eqnarray}
\Pi\;^{^{p=\frac{1}{2}}}(q)&:=&\frac{1}{\sqrt{1+(\frac{q}{\beta\epsilon_{0}A})^{2}}},\\
\Pi\;^{^{p=\frac{2}{3}}}(q)&:=&\sqrt[3]{\frac{1}{2}-\frac{1}{27\beta^{6}}\bigg(\frac{q}{\epsilon_{0}A}\bigg)^{6}+\sqrt{\frac{1}{4}-\frac{1}{27\beta^{6}}\bigg(\frac{q}{\epsilon_{0}A}\bigg)^{6}}}\nonumber \\
\qquad &+&\sqrt[3]{\frac{1}{2}-\frac{1}{27\beta^{6}}\bigg(\frac{q}{\epsilon_{0}A}\bigg)^{6}-\sqrt{\frac{1}{4}-\frac{1}{27\beta^{6}}\bigg(\frac{q}{\epsilon_{0}A}\bigg)^{6}}}-\frac{1}{3\beta^{2}}\bigg(\frac{q}{\epsilon_{0}A}\bigg)^{2},\\
\Pi\;^{^{p=\frac{3}{4}}}(q)&:=&\sqrt{\sqrt{1+\frac{1}{4\beta^{4}}\bigg(\frac{q}{\epsilon_{0}A}\bigg)^{4}}-\frac{1}{2\beta^{2}}\bigg(\frac{q}{\epsilon_{0}A}\bigg)^{2}},\\
\Pi\;^{^{p=\frac{5}{6}}}(q)&:=&\sqrt{\sqrt[3]{\frac{1}{2}+\sqrt{\frac{1}{4}+\frac{1}{27\beta^{6}}\bigg(\frac{q}{\epsilon_{0}A}\bigg)^{6}}}+\sqrt[3]{\frac{1}{2}-\sqrt{\frac{1}{4}+\frac{1}{27\beta^{6}}\bigg(\frac{q}{\epsilon_{0}A}\bigg)^{6}}}}.
\end{eqnarray}
\end{subequations}
When the Born-Infeld parameter $\beta$ takes the large values, the
behavior of the electric fields in (35a)-(35d) are given by
\begin{subequations}
\begin{eqnarray}
E_{z}\;^{^{p=\frac{1}{2}}}&=&\bigg(\frac{q}{\epsilon_{0}A}\bigg)\bigg[1-\frac{1}{2}\beta^{-2}\bigg(\frac{q}{\epsilon_{0}A}\bigg)^{2}+\frac{3}{8}\beta^{-4}\bigg(\frac{q}{\epsilon_{0}A}\bigg)^{4}+{\cal
O}\big(\beta^{-6}\big)\bigg],\\
E_{z}\;^{^{p=\frac{2}{3}}}&=&\bigg(\frac{q}{\epsilon_{0}A}\bigg)\bigg[1-\frac{1}{3}\beta^{-2}\bigg(\frac{q}{\epsilon_{0}A}\bigg)^{2}+\frac{1}{9}\beta^{-4}\bigg(\frac{q}{\epsilon_{0}A}\bigg)^{4}+{\cal
O}\big(\beta^{-6}\big)\bigg],\\
E_{z}\;^{^{p=\frac{3}{4}}}&=&\bigg(\frac{q}{\epsilon_{0}A}\bigg)\bigg[1-\frac{1}{4}\beta^{-2}\bigg(\frac{q}{\epsilon_{0}A}\bigg)^{2}+\frac{1}{32}\beta^{-4}\bigg(\frac{q}{\epsilon_{0}A}\bigg)^{4}+{\cal
O}\big(\beta^{-6}\big)\bigg],\\
E_{z}\;^{^{p=\frac{5}{6}}}&=&\bigg(\frac{q}{\epsilon_{0}A}\bigg)\bigg[1-\frac{1}{6}\beta^{-2}\bigg(\frac{q}{\epsilon_{0}A}\bigg)^{2}-\frac{1}{72}\beta^{-4}\bigg(\frac{q}{\epsilon_{0}A}\bigg)^{4}+{\cal
O}\big(\beta^{-6}\big)\bigg].
\end{eqnarray}
\end{subequations}
It must be emphasized that in obtaining the above results, the
Mathematica software has been used [27]. All of the electric
fields (37a)-(37d), have the Maxwellian limit in the weak field
regime. The first term on the right-hand side of (37a)-(37d)
represent the electric field between the plates of a
parallel-plate capacitor in Maxwell theory, while the higher-order
terms represent the effect of nonlinear corrections. Using (19),
it is obvious that we can write $\textbf{E}(\textbf{x})$ in the
following way:
\begin{equation}
\textbf{E}(\textbf{x})=-\boldsymbol{\nabla}\phi(\textbf{x}),
\end{equation}
where $\phi(\textbf{x})$ is the electrostatic potential. From (38)
we obtain the following relation:
\begin{equation}
\triangle\phi=-\int^{f}_{i}\textbf{E}(\textbf{x}).d\textbf{l},
\end{equation}
where $\triangle\phi=\phi_{f}-\phi_{i}$ is the potential
difference between the initial and final points, and $d\textbf{l}$
is an infinitesimal displacement vector. If we use Eqs. (35) and
(39), we will get the following expressions for the potential
difference between the plates of a nonlinear parallel-plate
capacitor for
$p\in\lbrace\frac{1}{2},\frac{2}{3},\frac{3}{4},\frac{5}{6}\rbrace$
\begin{subequations}
\begin{eqnarray}
\triangle\phi\;^{^{p=\frac{1}{2}}}&=&-\int_{-}^{+}\frac{q}{\epsilon_{0}A}\Pi\;^{^{p=\frac{1}{2}}}(q)(-\,\hat{\textbf{e}}_{z}).(\hat{\textbf{e}}_{z}\,
dz)\nonumber \\
\qquad &=&\frac{q d}{\epsilon_{0}A}\Pi\;^{^{p=\frac{1}{2}}}(q),\\
\triangle\phi\;^{^{p=\frac{2}{3}}}&=&-\int_{-}^{+}\frac{q}{\epsilon_{0}A}\Pi\;^{^{p=\frac{2}{3}}}(q)(-\,\hat{\textbf{e}}_{z}).(\hat{\textbf{e}}_{z}\,
dz)\nonumber \\
\qquad &=&\frac{q d}{\epsilon_{0}A}\Pi\;^{^{p=\frac{2}{3}}}(q),\\
\triangle\phi\;^{^{p=\frac{3}{4}}}&=&-\int_{-}^{+}\frac{q}{\epsilon_{0}A}\Pi\;^{^{p=\frac{3}{4}}}(q)(-\,\hat{\textbf{e}}_{z}).(\hat{\textbf{e}}_{z}\,
dz)\nonumber \\
\qquad &=&\frac{q d}{\epsilon_{0}A}\Pi\;^{^{p=\frac{3}{4}}}(q),\\
\triangle\phi\;^{^{p=\frac{5}{6}}}&=&-\int_{-}^{+}\frac{q}{\epsilon_{0}A}\Pi\;^{^{p=\frac{5}{6}}}(q)(-\,\hat{\textbf{e}}_{z}).(\hat{\textbf{e}}_{z}\,
dz)\nonumber \\
\qquad &=&\frac{q d}{\epsilon_{0}A}\Pi\;^{^{p=\frac{5}{6}}}(q).
\end{eqnarray}
\end{subequations}
As is well known, in electrostatics the capacitance $C$ of a
capacitor is the ratio of the amount of charge on each plate of a
capacitor to the potential difference between the plates of the
capacitor, i.e.,
\begin{equation}
C=\frac{q}{\triangle\phi}.
\end{equation}
It is necessary to note that Eq. (41) is also applicable for
determination of the capacitance of nonlinear capacitors [2,17,18].
After inserting (40) into (41), the capacitance of a nonlinear
parallel-plate capacitor in modified Born-Infeld electrostatics for
$p\in\lbrace\frac{1}{2},\frac{2}{3},\frac{3}{4},\frac{5}{6}\rbrace$
becomes
\begin{subequations}
\begin{eqnarray}
C\;^{^{p=\frac{1}{2}}}&=&\frac{\epsilon_{0}A}{d}\frac{1}{\Pi\;^{^{p=\frac{1}{2}}}(q)},\\
C\;^{^{p=\frac{2}{3}}}&=&\frac{\epsilon_{0}A}{d}\frac{1}{\Pi\;^{^{p=\frac{2}{3}}}(q)},\\
C\;^{^{p=\frac{3}{4}}}&=&\frac{\epsilon_{0}A}{d}\frac{1}{\Pi\;^{^{p=\frac{3}{4}}}(q)},\\
C\;^{^{p=\frac{5}{6}}}&=&\frac{\epsilon_{0}A}{d}\frac{1}{\Pi\;^{^{p=\frac{5}{6}}}(q)}.
\end{eqnarray}
The above equations have the following weak field expansions:
\begin{eqnarray}
C\;^{^{p=\frac{1}{2}}}&=&\bigg(\frac{\epsilon_{0}A}{d}\bigg)\bigg[1+\frac{1}{2}\beta^{-2}\bigg(\frac{q}{\epsilon_{0}A}\bigg)^{2}-\frac{1}{8}\beta^{-4}\bigg(\frac{q}{\epsilon_{0}A}\bigg)^{4}+{\cal
O}\big(\beta^{-6}\big)\bigg],\\
C\;^{^{p=\frac{2}{3}}}&=&\bigg(\frac{\epsilon_{0}A}{d}\bigg)\bigg[1+\frac{1}{3}\beta^{-2}\bigg(\frac{q}{\epsilon_{0}A}\bigg)^{2}-\frac{1}{81}\beta^{-6}\bigg(\frac{q}{\epsilon_{0}A}\bigg)^{6}+{\cal
O}\big(\beta^{-8}\big)\bigg],\\
C\;^{^{p=\frac{3}{4}}}&=&\bigg(\frac{\epsilon_{0}A}{d}\bigg)\bigg[1+\frac{1}{4}\beta^{-2}\bigg(\frac{q}{\epsilon_{0}A}\bigg)^{2}+\frac{1}{32}\beta^{-4}\bigg(\frac{q}{\epsilon_{0}A}\bigg)^{4}+{\cal
O}\big(\beta^{-6}\big)\bigg],\\
C\;^{^{p=\frac{5}{6}}}&=&\bigg(\frac{\epsilon_{0}A}{d}\bigg)\bigg[1+\frac{1}{6}\beta^{-2}\bigg(\frac{q}{\epsilon_{0}A}\bigg)^{2}+\frac{1}{24}\beta^{-4}\bigg(\frac{q}{\epsilon_{0}A}\bigg)^{4}+{\cal
O}\big(\beta^{-6}\big)\bigg].
\end{eqnarray}
\end{subequations}
Equations (42e)-(42h) show that the capacitance of a nonlinear
parallel-plate capacitor in modified Born-Infeld electrostatics is a
function of the amount of charge on each plate of the capacitor.
Now, let us compute the energy density between the plates of a
nonlinear parallel-plate capacitor in modified Born-Infeld
electrostatics. If we put (35) into (32), we will get the following
results:
\begin{subequations}
\begin{eqnarray}
u(\textbf{x})\;^{^{p=\frac{1}{2}}}&=&\epsilon_{0}\beta^{2}\bigg[\frac{1}{\Pi\;^{^{p=\frac{1}{2}}}(q)}-1\bigg], \\
u(\textbf{x})\;^{^{p=\frac{2}{3}}}&=&\frac{3}{4}\epsilon_{0}\beta^{2}\bigg[\frac{\frac{1}{3\beta^{2}}\bigg(\frac{q}{\epsilon_{0}A}\Pi\;^{^{p=\frac{2}{3}}}(q)\bigg)^{2}+1}{\Pi\;^{^{p=\frac{2}{3}}}(q)}-1\bigg],\\
u(\textbf{x})\;^{^{p=\frac{3}{4}}}&=&\frac{2}{3}\epsilon_{0}\beta^{2}\bigg[\frac{\frac{1}{2\beta^{2}}\bigg(\frac{q}{\epsilon_{0}A}\Pi\;^{^{p=\frac{3}{4}}}(q)\bigg)^{2}+1}{\Pi\;^{^{p=\frac{3}{4}}}(q)}-1\bigg],\\
u(\textbf{x})\;^{^{p=\frac{5}{6}}}&=&\frac{3}{5}\epsilon_{0}\beta^{2}\bigg[\frac{\frac{2}{3\beta^{2}}\bigg(\frac{q}{\epsilon_{0}A}\Pi\;^{^{p=\frac{5}{6}}}(q)\bigg)^{2}+1}{\Pi\;^{^{p=\frac{5}{6}}}(q)}-1\bigg].
\end{eqnarray}
\end{subequations}
Using Eq. (43), the electrostatic potential energy of a nonlinear
parallel-plate capacitor in modified Born-Infeld electrostatics
according to Figure 1 becomes
\begin{subequations}
\begin{eqnarray}
U\;^{^{p=\frac{1}{2}}}&=&\int_{{area\;of\;a\;plate}}da\;\int_{0}^{d}dz\, u(\textbf{x})\;^{^{p=\frac{1}{2}}}\nonumber \\
\qquad&=&\epsilon_{0}\beta^{2}\bigg[\sqrt{1+(\frac{q}{\beta\epsilon_{0}A})^{2}}-1\bigg]Ad,\\
U\;^{^{p=\frac{2}{3}}}&=&\int_{{area\;of\;a\;plate}}da\;\int_{0}^{d}dz\, u(\textbf{x})\;^{^{p=\frac{2}{3}}}\nonumber \\
\qquad&=&\frac{3}{4}\epsilon_{0}\beta^{2}\bigg[\frac{\frac{1}{3\beta^{2}}\bigg(\frac{q}{\epsilon_{0}A}\Pi\;^{^{p=\frac{2}{3}}}(q)\bigg)^{2}+1}{\Pi\;^{^{p=\frac{2}{3}}}(q)}-1\bigg]Ad,\\
U\;^{^{p=\frac{3}{4}}}&=&\int_{{area\;of\;a\;plate}}da\;\int_{0}^{d}dz\, u(\textbf{x})\;^{^{p=\frac{3}{4}}}\nonumber \\
\qquad&=&\frac{2}{3}\epsilon_{0}\beta^{2}\bigg[\frac{\frac{1}{2\beta^{2}}\bigg(\frac{q}{\epsilon_{0}A}\Pi\;^{^{p=\frac{3}{4}}}(q)\bigg)^{2}+1}{\Pi\;^{^{p=\frac{3}{4}}}(q)}-1\bigg]Ad,\\
U\;^{^{p=\frac{5}{6}}}&=&\int_{{area\;of\;a\;plate}}da\;\int_{0}^{d}dz\, u(\textbf{x})\;^{^{p=\frac{5}{6}}}\nonumber \\
\qquad&=&\frac{3}{5}\epsilon_{0}\beta^{2}\bigg[\frac{\frac{2}{3\beta^{2}}\bigg(\frac{q}{\epsilon_{0}A}\Pi\;^{^{p=\frac{5}{6}}}(q)\bigg)^{2}+1}{\Pi\;^{^{p=\frac{5}{6}}}(q)}-1\bigg]Ad.
\end{eqnarray}
\end{subequations}
For the large values of the Born-Infeld parameter $\beta$, the
behavior of the electrostatic potential energies in (44a)-(44d)
are given by
\begin{subequations}
\begin{eqnarray}
U\;^{^{p=\frac{1}{2}}}|_{_{large \;
\beta}}&=&U_{_{M}}\bigg[1-\frac{1}{4}\beta^{-2}\bigg(\frac{q}{\epsilon_{0}A}\bigg)^{2}+\frac{1}{8}\beta^{-4}\bigg(\frac{q}{\epsilon_{0}A}\bigg)^{4}+{\cal
O}\big(\beta^{-6}\big)\bigg],\\
U\;^{^{p=\frac{2}{3}}}|_{_{large \;
\beta}}&=&U_{_{M}}\bigg[1-\frac{1}{6}\beta^{-2}\bigg(\frac{q}{\epsilon_{0}A}\bigg)^{2}+\frac{1}{27}\beta^{-4}\bigg(\frac{q}{\epsilon_{0}A}\bigg)^{4}+{\cal
O}\big(\beta^{-6}\big)\bigg],\\
U\;^{^{p=\frac{3}{4}}}|_{_{large \;
\beta}}&=&U_{_{M}}\bigg[1-\frac{1}{8}\beta^{-2}\bigg(\frac{q}{\epsilon_{0}A}\bigg)^{2}+\frac{1}{96}\beta^{-4}\bigg(\frac{q}{\epsilon_{0}A}\bigg)^{4}+{\cal
O}\big(\beta^{-6}\big)\bigg],\\
U\;^{^{p=\frac{5}{6}}}|_{_{large \;
\beta}}&=&U_{_{M}}\bigg[1-\frac{1}{12}\beta^{-2}\bigg(\frac{q}{\epsilon_{0}A}\bigg)^{2}-\frac{1}{216}\beta^{-4}\bigg(\frac{q}{\epsilon_{0}A}\bigg)^{4}+{\cal
O}\big(\beta^{-6}\big)\bigg],
\end{eqnarray}
\end{subequations}
where $U_{_{M}}=\frac{q^{2}}{2C_{_{M}}}$ and
$C_{_{M}}=\frac{\epsilon_{0}A}{d}$ are the electrostatic potential
energy and the capacitance of a parallel-plate capacitor in
Maxwell electrostatics respectively. Equation (45) shows that the
relation $U_{_{M}}=\frac{q^{2}}{2C_{_{M}}}$ is not true for a
parallel-plate capacitor in modified Born-Infeld electrostatics.
In the limit of $\beta\rightarrow\infty$, equations (45a)-(45d)
reduce to the following equation:
\begin{equation}
U\;^{^{p=\frac{1}{2}}}|_{_{\beta=\infty}}=U\;^{^{p=\frac{2}{3}}}|_{_{\beta=\infty}}=U\;^{^{p=\frac{3}{4}}}|_{_{\beta=\infty}}=U\;^{^{p=\frac{5}{6}}}|_{_{\beta=\infty}}=U_{_{M}}.
\end{equation}

\section{Summary and Conclusions}
In 1930s Born-Infeld theory was introduced in order to remove the
infinite self-energy of the electron in Maxwell electrodynamics
[9]. In Born-Infeld electrodynamics the absolute value of the
electric field has an upper limit $\beta$, i.e., $|\textbf
E|\leq\beta$. In Born-Infeld paper the numerical value of $\beta$
was [9,28]:
\begin{equation}
\beta_{_{Born-Infeld}}=1.2\times10^{20}\frac{V}{m}.
\end{equation}
Soff, Rafelski, and Greiner obtained the following lower bound on
$\beta$ [29]:
\begin{equation}
\beta_{_{Soff}}\geq 1.7\times10^{22}\frac{V}{m}.
\end{equation}
In a paper about photon-photon scattering and photon splitting in
a magnetic field in Born-Infeld theory, Davila and his coworkers
obtained the following new lower bound on $\beta$ [30]:
\begin{equation}
\beta_{_{Davila}}\geq 2.0\times10^{19}\frac{V}{m}.
\end{equation}
In 1983, E. Iacopini and E. Zavattini introduced a
$(p,\tau)$-two-parameter modification of Born-Infeld
electrodynamics, in which the self-energy of a point-like charge
becomes finite for $p<1$ [19]. In our paper, after a brief
formulation of Born-Infeld-Iacopini-Zavattini electrodynamics
(modified Born-Infeld electrodynamics) in the presence of an
external current, the capacitance and the electrostatic potential
energy of a nonlinear parallel-plate capacitor have been calculated
exactly in the framework of modified Born-Infeld electrostatics for
$p\in\lbrace\frac{1}{2},\frac{2}{3},\frac{3}{4},\frac{5}{6}\rbrace$.
In order to have a deeper understanding of nonlinear effects in
modified Born-Infeld electrostatics, let us rewrite (45a) as
follows:
\begin{equation}
U\;^{^{p=\frac{1}{2}}}|_{_{large \; \beta}}=U_{_{M}}
+U\;^{^{first-order\;nonlinear\;correction}}_{_{M}}+U\;^{^{second-order\;nonlinear\;correction}}_{_{M}}+{\cal
O}\big(\beta^{-6}\big),
\end{equation}
where
\begin{subequations}
\begin{eqnarray}
U\;^{^{first-order\;nonlinear\;correction}}_{_{M}}&:=&-\frac{1}{2\beta^{2}\epsilon_{0}Ad}U_{_{M}}\;^{^{2}},\\
U\;^{^{second-order\;nonlinear\;correction}}_{_{M}}&:=&\frac{1}{2\beta^{4}\epsilon_{0}^{2}A^{2}d^{2}}U_{_{M}}\;^{^{3}}.
\end{eqnarray}
\end{subequations}
Now, let us estimate the numerical values of $U_{_{M}}$,
$U\;^{^{first-order\;nonlinear\;correction}}_{_{M}}$, and
$U\;^{^{second-order\;nonlinear\;correction}}_{_{M}}$ in (50). For
this aim, we use the following typical values for a parallel-plate
capacitor (see page 804 in Ref. [31]):
\begin{equation}
A=100 \;cm^{2},\; d=1.0 \;mm,\; q=1.06\times10^{-9} \;C.
\end{equation}
If we put Eqs. (47), (48), (49), and (52) into Eq. (50), we get
\begin{subequations}
\begin{eqnarray}
U_{_{M}}&=&6.35\times10^{-9}\;J,\\
U\;^{^{first-order\;nonlinear\;correction}}_{_{M\;(Born-Infeld)}}&=&-1.58\times10^{-41}\;J,\\
U\;^{^{second-order\;nonlinear\;correction}}_{_{M\;(Born-Infeld)}}&=&7.88\times10^{-74}\;J,\\
U\;^{^{first-order\;nonlinear\;correction}}_{_{M\;(Soff)}}&=&-7.88\times10^{-46}\;J,\\
U\;^{^{second-order\;nonlinear\;correction}}_{_{M\;(Soff)}}&=&1.96\times10^{-82}\;J,\\
U\;^{^{first-order\;nonlinear\;correction}}_{_{M\;(Davila)}}&=&-5.69\times10^{-40}\;J,\\
U\;^{^{second-order\;nonlinear\;correction}}_{_{M\;(Davila)}}&=&1.02\times10^{-70}\;J.
\end{eqnarray}
\end{subequations}
It  must be noted that in (53d)-(53g) the minimum value of $\beta$
in Eqs. (48) and (49) has been used. Equations (53a)-(53g) tell us
that the nonlinear corrections to electrostatic potential energy
in a parallel-plate capacitor are not important in the weak field
limit. For $p=\frac{7}{8}(\Gamma=8)$, Eq. (34) becomes
\begin{equation}
E_{z}\;^{^{8}}+\bigg(\frac{q}{\beta^{\frac{1}{4}}\epsilon_{0}A}
\bigg)^{^{8}}\; E_{z}\;^{^{2}}-\bigg(\frac{q}{\epsilon_{0}A}
\bigg)^{^{8}}=0.
\end{equation}
Equation (54) is an eight-order equation which by a suitable change
of variable reduces to a quartic equation. In future studies we want
to obtain the exact solutions of (54) in order to calculate the
capacitance and the electrostatic potential energy of a nonlinear
parallel-plate capacitor in modified Born-Infeld electrostatics for
$p=\frac{7}{8}$. More recently, a new modification of Born-Infeld
electrodynamics has been developed which includes three independent
parameters [32]. We hope to study the problems discussed in our work
from the viewpoint of [32] in an independent research.




\begin{thebibliography}{11}
\bibitem{P1}
N. S. Kuek, A. C. Liew, E. Schamiloglu, and J. O. Rossi, IEEE
Transactions on Plasma Science \textbf{40}, 2523 (2012).
\bibitem{P2}
R. E. Vermillion, Eur. J. Phys. \textbf{19}, 173 (1998).
\bibitem{P3}
E. Gluskin, J. Franklin Inst. \textbf{336}, 1035 (1999).
\bibitem{P4}
E. Gluskin, Int. J. Electron. \textbf{58}, 63 (1985).
\bibitem{P5}
F. Tao, W. Chen, W. Xu, J. Pan, and S. Du, Phys. Rev. E
\textbf{83}, 056605 (2011).
\bibitem{P6}
A. I. Khan, D. Bhowmik, P. Yu, S. J. Kim, X. Pan, R. Ramesh, and
S. Salahuddin, Appl. Phys. Lett. \textbf{99}, 113501 (2011).
\bibitem{P7}
A. F. Kabychenkov and F. V. Lisovskii, J. Exp. Theor. Phys.
\textbf{118}, 643 (2014).
\bibitem{P8}
J. D. Jackson, Classical Electrodynamics, 3rd edition (John Wiley,
1999).
\bibitem{P9}
M. Born and L. Infeld, Proc. R. Soc. London A \textbf{144}, 425
(1934).
\bibitem{P10}
B. Zwiebach, A First Course in String Theory, 2nd edition
(Cambridge University Press, Cambridge, UK, 2009).
\bibitem{P11}
S. K. Moayedi, M. Shafabakhsh, and F. Fathi, Adv. High Energy
Phys. \textbf{2015}, 180185 (2015).
\bibitem{P12}
H. Euler and B. Kockel, Naturwissenschaften \textbf{23}, 246
(1935).
\bibitem{P13}
H. Euler, Ann. Phys. (Leipzig) \textbf{418}, 398 (1936).
\bibitem{P14}
W. Heisenberg and H. Euler, Zeitschrift fur Physik \textbf{98},
714 (1936).
\bibitem{P15}
R. Battesti and C. Rizzo, Rep. Prog. Phys. \textbf{76}, 016401
(2013).
\bibitem{P16}
M. Fouche, R. Battesti, and C. Rizzo, Phys. Rev. D \textbf{93},
093020 (2016).
\bibitem{P17}
G. Munoz, Am. J. Phys. \textbf{64}, 1285 (1996).
\bibitem{P18}
S. K. Moayedi and M. Shafabakhsh, Eur. Phys. J. Plus \textbf{131},
55 (2016).
\bibitem{P19}
E. Iacopini and E. Zavattini, Nuovo Cimento B \textbf{78}, 38
(1983).
\bibitem{P20}
R. A. Bertlmann, Anomalies in Quantum Field Theory (Oxford Science
Publications, Oxford, 2000).
\bibitem{P21}
F. J. Belinfante, Physica \textbf{6}, 887 (1939).
\bibitem{P22}
M. Montesinos and E. Flores, Rev. Mex. Fis. \textbf{52}, 29
(2006).
\bibitem{P23}
G. Munoz, Am. J. Phys. \textbf{64}, 1153 (1996).
\bibitem{P24}
A. Accioly,  Am. J. Phys. \textbf{65}, 882 (1997).
\bibitem{P25}
J.-P. Tignol, Galois' Theory of Algebraic Equations (World
Scientific, 2001).
\bibitem{P26}
S. C. Newman, A Classical Introduction to Galois Theory (Wiley,
2012).
\bibitem{P27}
A. Grozin, Introduction to Mathematica for Physicists (Springer,
2014).
\bibitem{P28}
G. Boillat and A. Strumia, J. Math. Phys. \textbf{40}, 1 (1999).
\bibitem{P29}
G. Soff, J. Rafelski, and W. Greiner, Phys. Rev. A \textbf{7}, 903
(1973).
\bibitem{P30}
J. M. Davila, C. Schubert, and M. A. Trejo, Int. J. Mod. Phys. A
\textbf{29}, 1450174 (2014).
\bibitem{P31}
P. A. Tipler and G. Mosca, Physics for Scientists and Engineers,
6th edition (Freeman, New York, 2008).
\bibitem{P32}
S. I. Kruglov, ``Notes on Born-Infeld-type electrodynamics'',
arXiv:1612.04195v2.


\end{thebibliography}
\end{document}